\documentclass[aps,prl,twocolumn,superscriptaddress,showpacs,floatfix,amsmath,amssymb]{revtex4-1}
\usepackage{amssymb}
\usepackage{bm}
\usepackage{graphicx}

\usepackage[dvipsnames,usenames]{color}
\begin{document}

\title{Clustering of gyrotactic microorganisms in turbulent flows}

\author{Filippo De Lillo} \affiliation{Dipartimento di Ingegneria delle Costruzioni,
  dell'Ambiente e del Territorio, Universit\`a di Genova, via
  Montallegro 1, 16145 Genova, Italy}\affiliation{Dipartimento di Fisica 
  and INFN, Universit\`a di Torino, via P. Giuria 1, 10125 Torino,
  Italy}

\author{Guido Boffetta}\affiliation{Dipartimento di Fisica 
  and INFN, Universit\`a di Torino, via P. Giuria 1, 10125 Torino,
  Italy}

\author{Massimo Cencini}\affiliation{Istituto dei Sistemi Complessi,
  Consiglio Nazionale delle Ricerche, via dei Taurini 19, 00185
  Rome, Italy}
\begin{abstract}
  We study the spatial distribution of gyrotactic microorganisms
  transported by a three-dimensional turbulent flow generated by
  direct numerical simulations. We find that gyrotaxis combines with
  turbulent fluctuations to produce small scales (multi-)fractal
  clustering. We explain this result by showing that gyrotactic
  swimming cells behave like tracers in a compressible flow.  The
  effective compressibility is derived in the limits of fluid
  acceleration much larger and smaller than the gravity.
\end{abstract}
\maketitle 

Microbial patchiness in oceans is important for ecological and
evolutionary dynamics \cite{Levin1994,Azam2007} and for biogeochemical
processes \cite{Falkowski2000}.  In motile aquatic microorganisms,
self-propulsion provides a mechanism to escape fluid pathlines,
potentially leading to small-scale patchiness
\cite{Mitchell1990,Torney2007,Durham2011}.  Remarkably, motility
combined with fluid flows can also generate large-scale
inhomogeneities. For instance, spectacular aggregation of
phytoplankton cells (in layers centimeters to meters thin,
horizontally extending from hundreds of meters to kilometers) can
result from vertical shears and \textit{gyrotactic swimming}
\cite{Durham2009}.  Gyrotaxis characterizes several species of motile
microalgae whose swimming direction is determined by the balance of
viscous and gravitational torques, due to the displacement between the
cell center of mass and buoyancy. As an effect of such balance, for
example, gyrotactic algae aggregate in the center (wall) of descending
(ascending) vertical pipe flows \cite{Kessler1985,Pedley1987}.
Gyrotaxis is observed in algae, e.g., of the genus
\textit{Chlamydomonas}, which can be engineered to transport
microloads \cite{Weibel2005}, or \textit{Dunaliella}, employed in
biofuels \cite{Chisti2007}.  So far most studies focused on the
dynamics of gyrotactic microorganisms in simple stationary flows or
kinematic models
\cite{Kessler1985,Pedley1987,Hill2002,Durham2009,Thorn2010,Durham2011}.

\begin{figure}[t!]
\centering
{\includegraphics[width=0.47\columnwidth]{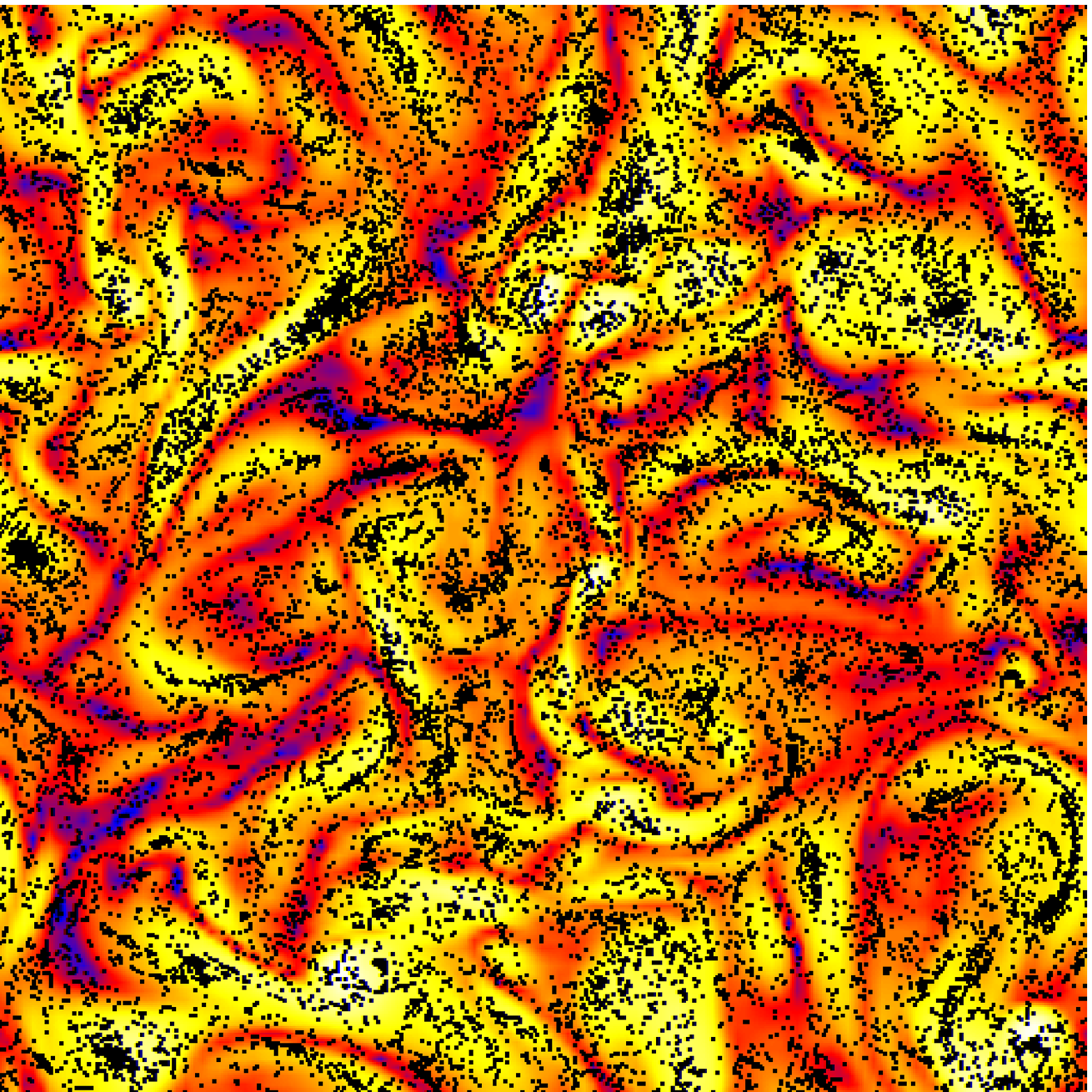}
\hfill \includegraphics[width=0.47\columnwidth]{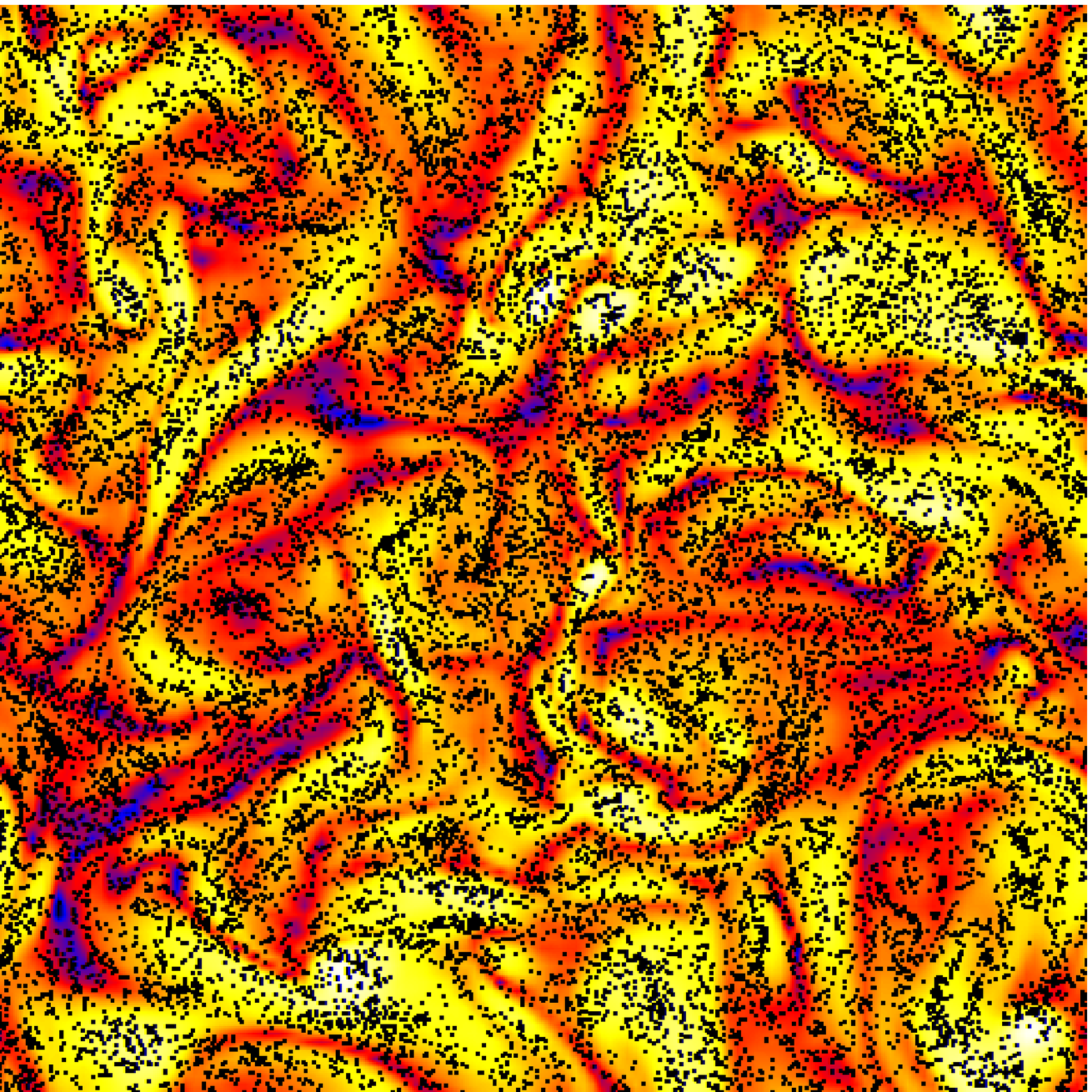}}
\caption{(color online) Spatial distribution of gyrotactic swimmers (dots) in
  a slab of a 3D turbulent flow.  Color code: yellow/blue corresponds
  to high/low vorticity values ($\ln |\bm
  \omega|/\omega_{\mathrm{rms}}$).  (Left) Limit of orientation
  dominated by local fluid acceleration ($\bm A=\bm a$, see text) with
  aggregation in high vorticity regions.  (Right) Limit of gravity
  dominated orientation ($\bm A=-\bm g$). Parameters correspond to
  circled symbols in Fig.~\ref{fig2}c and Fig.~\ref{fig3}a, respectively.}
\label{fig1}
\end{figure}

In this Letter, we investigate the interplay between gyrotactic
motility and realistic turbulent flows, as occurring in the sea.  We
find that turbulence and gyrotaxis combine to generate inhomogeneous
distributions with small-scale (multi-)fractal statistics (see
Fig.~\ref{fig1}).  We study the limit of gravitational acceleration
much smaller or larger than turbulent accelerations to identify the
mechanisms responsible for gyrotactic clustering in terms of
an effective compressible velocity field.

We consider dilute suspensions of non interacting motile
microorganisms, much smaller than the smallest scale of turbulence,
the Kolmogorov length $\eta$.  We can thus model them as
self-propelled particles with velocity,
\begin{equation}
\dot{\bm X} = {\bm u}({\bm X},t) + v_s {\mathrm{\bf p}}\,,
\label{eq:1}
\end{equation}
given by the sum of the fluid velocity $\bm u$ at the particle
position $\bm X$ and the swimming contribution $v_s {\mathrm{\bf p}}$,
where the swimming speed $v_s$ is assumed constant
\cite{Pedley1987,Torney2007}. Cells are assumed spherical and
neutrally buoyant, with the center of mass displaced by $h$ with
respect to the geometric one. The swimming direction ${\mathrm{\bf
    p}}$, determined by the total torque acting on the cell, evolves
as
\begin{equation}
\dot{{\mathrm{\bf p}}} = \frac{1}{2v_o}\left[{\bm A} - ({\bm A} \cdot {\mathrm{\bf p}})
{\mathrm{\bf p}} \right] +\frac{1}{2} {\bm \omega} \times {\mathrm{\bf p}}\,,
\label{eq:2}
\end{equation}
where ${\bm \omega}$ is the fluid vorticity and $v_o=3 \nu/h$ is the
orientation speed for spherical cells subject to the acceleration
${\bm A}$ \cite{Pedley1987}.  In a fluid at rest, besides viscous
forces, only gravity (and buoyancy) $\bm g$ is acting
and thus ${\bm A}=-{\bm g}=g\hat{\bm z}$, while acceleration due to
swimming is neglected \cite{Pedley1987}.  In presence of a flow, we
have ${\bm A}={\bm a}-{\bm g}$ where
\begin{equation}
\bm a \equiv \partial_t{\bm u} + {\bm u} \cdot {\bm \nabla} {\bm u} = - {\bm
  \nabla} p + \nu \nabla^2 {\bm u} + {\bm f} \qquad 
\label{eq:3}
\end{equation}
is the fluid acceleration given by the Navier-Stokes equations ruling
the velocity $\bm u$ of an incompressible ($\bm \nabla \cdot \bm u=0$)
fluid with viscosity $\nu$, pressure $p$ and stirred by an external
forcing ${\bm f}$.  Previous studies on gyrotactic swimmers
disregarded fluid acceleration, as mainly focused on
simple, non-turbulent flows where $|\bm a| \ll g$. In turbulence,
fluid acceleration can locally exceed $g$ \cite{LaPorta2001} and
therefore its contribution has to be taken into account.

The first term on the rhs of Eq.~(\ref{eq:2}) causes the direction of
swimming $\mathrm{\bf p}$ to align with ${\bm A}$ on a time scale $v_o
/A$.  When the contribution of fluid acceleration can be neglected,
cells tend to orient vertically ($\mathrm{\bf p} \to \hat{\bm z}$) on
a time scale $B=v_o/g$.  The alignment is contrasted by the vorticity
term ${\bm \omega} \times \mathrm{\bf p}$ and, depending on $B\omega$
being smaller or larger than $1$, cells may swim along a resulting
local equilibrium direction or tumble randomly as the orientation
becomes unstable due to vorticity \cite{Pedley1987,Thorn2010}. In
principle, the swimming direction may be modified also by rotational
Brownian motion \cite{Berg1993} and tumbling due to flagella
desynchronization during swimming \cite{Polin2009}, which are here
neglected. The former effect is very small for typical algae (having
size $\mathcal{O}(10\mu m)$); the latter can be neglected whenever the
tumbling time is longer than the reorientation one.

We study gyrotactic swimming in homogeneous and isotropic turbulent
velocity fields of moderate intensity ($Re_\lambda \approx 65-100$) by
means of direct numerical simulations of Navier-Stokes
equations.  In particular, Eq.~(\ref{eq:3}) is solved by means of a
standard pseudospectral algorithm with 2$^{nd}$ order Runge-Kutta
time-stepping, on a tri-periodic cubic grid of size $N^3$ (for
$N\!=\!128$ and $256$). 
Statistical stationarity is guaranteed by means of a zero mean,
Gaussian and white in time random forcing ${\bm f}$ restricted to
large scales.  Viscosity $\nu$ is such that the Kolmogorov length
$\eta$ is of the order of the grid spacing, ensuring well resolved
small-scale velocity dynamics. 
For different values of $g$, 
several populations of swimmers,
characterized by different values of $v_s$ and $v_o$ are injected with
random positions and orientations.  At each time step, velocity and
acceleration at the swimmers positions, needed to integrate
Eqs.~(\ref{eq:1}-\ref{eq:2}), are obtained by interpolation.  The
self-propelled particles are then evolved, and their distribution and
orientation studied in statistically steady conditions. In the sequel,
we mostly focus on the dependence on the orientation speed $v_o$ by fixing
$v_s \approx 0.3 u_\eta$, $u_\eta$ being the typical fluid velocity
fluctuation at the Kolmogorov scale.
\begin{figure}[t!]
\centering
\includegraphics[width=0.9\columnwidth]{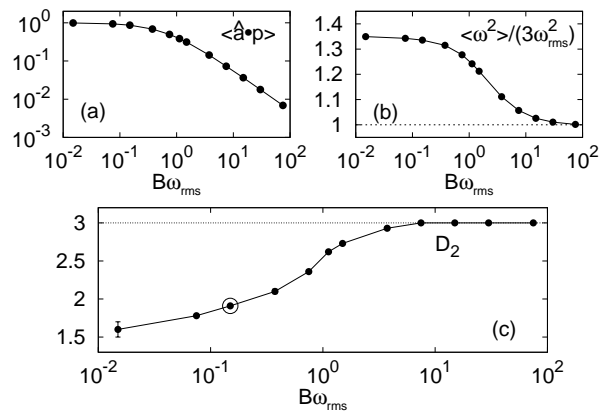}
\caption{Swimmer properties as a function of the orientation parameter
  $B \omega_{\mathrm{rms}}$ ($B=v_o/a_{\mathrm{rms}}$) 
in the limit $|\bm a| \gg g$.
(a) Average alignment with fluid acceleration 
$\langle \hat{\bm a} \cdot \mathrm{\bf p} \rangle$.
(b) Average square vorticity at swimmers position normalized to the 
volume average value.
(c) correlation dimension $D_2$. Circled symbol in (c) corresponds
to the distribution shown in Fig.~\ref{fig1}a.}
\label{fig2}
\end{figure}

Formally, Eqs.~(\ref{eq:1}-\ref{eq:2}) define a dissipative dynamical
system evolving in the $2d$-dimensional (actually $2d\!-\!1$ because
$\mathrm{p}^2\!=\!1$ and $d=3$) phase space ($\bm X$,$\mathrm{\bf p}$)
with phase-space contraction rate
\begin{equation}
\Gamma=\sum_{i=1}^{d} \frac{\partial \dot{X}_i}{\partial X_i}+
\sum_{i=1}^{d} \frac{\partial \dot{\mathrm{p}}_i}{\partial \mathrm{p}_i}=
-\frac{d+1}{2 v_o} 
\left(g \mathrm{p}_z + {\bm a} \cdot {\mathrm{\bf p}} \right)\,.
\label{eq:4}
\end{equation}
As $\mathrm{\bf p}$ orients in the direction $\bm a-\bm g$, $\Gamma$
is expected to be negative on average, meaning that swimmers will
evolve onto a dynamical attractor of dimension smaller than the 
whole phase space, which explains
why clustering can be observed: if the fractal dimension of
the attractor is smaller than $d$, clustering in position space (as in
Fig.~\ref{fig1}) is possible (see Ref.~\cite{Bec2005} for a
conceptually similar phenomenon occurring for inertial particles). 
We remark that clustering is a consequence of swimming: indeed for
$v_s=0$ Eqs.~(\ref{eq:1}) and (\ref{eq:2}) decouple, thus cells become
tracers advected by an incompressible velocity and cannot cluster.
Moreover, in the limit $v_o\to \infty$ we have $\Gamma \to 0$ and
therefore swimmers cannot cluster. Nonetheless, even in this limit,
if $v_s>0$ they deviate from fluid trajectories and generate 
interesting dynamics \cite{Ouellette2011}.

We now discuss the physical mechanisms of clustering which, as
anticipated, depend on whether the dominating effect comes from the
gravitational ($g$) or fluid acceleration (which we quantify in terms 
of its rms value $a_{\mathrm{rms}}$).

We start considering the case $a_{\mathrm{rms}} \gg\ g$ and therefore
we take $\bm A=\bm a$ in Eq.~(\ref{eq:2}).  Figure~\ref{fig2}
summarizes the behavior of the main observables as a function of the
dimensionless number $B \omega_{\mathrm{rms}}$ (now
$B=v_o/a_{\mathrm{rms}}$) measuring the ratio of the alignment
timescale to rotation timescale induced by vorticity.  When the
alignment is very fast, the swimming direction $\mathrm{\bf p}$
becomes parallel to the local direction of the fluid acceleration
$\hat{\bm a}={\bm a}/a$, as confirmed by Fig.~\ref{fig2}a showing that
$\langle \hat{\bm a} \cdot \mathrm{\bf p} \rangle \to 1$ for
$B\omega_{\mathrm{rms}}\ll 1$ (here and in the following
$\langle[\cdot]\rangle$ denotes average over particle
distribution). In this limit, swimming cells behave like tracers
advected by an effective velocity ${\bm v}\approx {\bm u}+v_s \hat{\bm
  a}$.  While $\bm u$ is incompressible, the effective velocity field
$\bm v$ is not: ${\bm \nabla} \cdot {\bm v} \propto v_s {\bm \nabla}
\cdot {\bm a}$ being negative (positive) in high vorticity (strain)
regions. Therefore, as it occurs for inertial particles lighter than
fluid \cite{Balkovsky2001,Calzavarini2008}, the swimmers cluster
inside vortical structures (Fig.~\ref{fig1}a and
Fig.~\ref{fig2}b). The divergence of $\bm v$ is proportional to $v_s$,
clustering is thus expected to increase with the swimming speed.  In
the opposite limit of slow alignment, when $B\omega_{\mathrm{rms}} \gg
1$, random tumbling due to fluid vorticity dominates, hence swimming
orientation cannot align to the local acceleration ($\langle \hat{\bm
  a} \cdot \mathrm{\bf p}\rangle\to 0$, see Fig.~\ref{fig2}a): the
compressible effect is lost and particles distribute uniformly in the
volume. To quantify clustering we measured the correlation dimension
$D_2$, ruling the small-distance ($r \to 0$) behavior of the
probability to find two swimmers at separation less than $r$:
$P_{2}(|\bm X_1 -\bm X_2|<r)\propto r^{D_2}$ \cite{paladin1987}.  For
uniformly distributed particles $D_2=d$, while when clustering is
present the probability to find close pairs increases and $D_2<d$ (see
e.g. \cite{Bec2007} for a similar study in the case of inertial
particles).  In Fig.~\ref{fig2}c we show $D_2$ as a function of
$B\omega_{\mathrm{rms}}$: for $B\omega_{\mathrm{rms}}\ll 1$,
$D_2\approx 1.5$, indicating strong clustering in almost filamental
structures; conversely, when $B\omega_{\mathrm{rms}} > 1$, the
correlation dimension approaches the uniform-distribution value
$D_2\approx 3$.
\begin{figure}[b!]
\centering
\includegraphics[width=1\columnwidth]{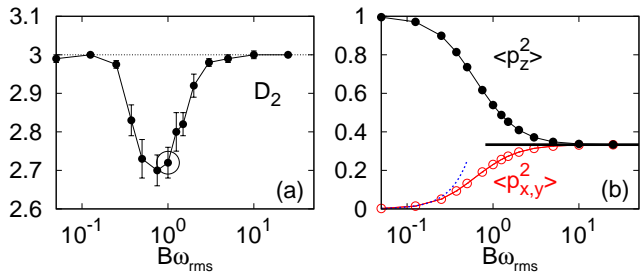}
\caption{(color online) Clustering properties as a function of 
$B\omega_{\mathrm{rms}}$, for $g \gg |\bm a|$ ($B=v_0/g$).
(a) Correlation dimension $D_2$ of the swimmer positions. Circled 
symbol corresponds to the data shown in Fig.~\ref{fig1}b.
(b) Variances of swimming direction components ($\langle
\mathrm{p}^2_{x}\rangle=\langle \mathrm{p}^2_{y}\rangle$ in red, and
$\langle \mathrm{p}^2_{z}\rangle$). The dashed blue curve is the
parabola $(B\omega_{\mathrm{rms}})^2$. The solid horizontal line
represents the random orientation value $1/3$.}
\label{fig3}
\end{figure}

\begin{figure}[t!]
\centering
\includegraphics[width=0.9\columnwidth]{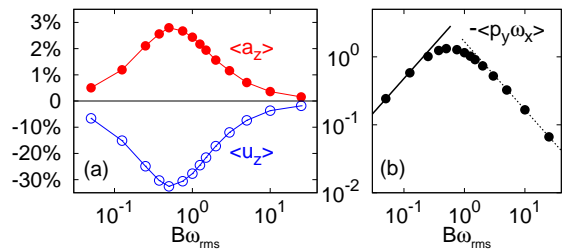}
\caption{(color online) (a) Average fluid acceleration (positive, red
  filled circles) and velocity (negative, blue open circles) along the
  vertical at swimmer positions, expressed in percentage of
  $a_{\mathrm{rms}}$ and $v_s$, respectively.  (b) Correlation
  $-\langle \mathrm{p}_y \omega_x\rangle$ vs
  $B\omega_{\mathrm{rms}}$. The maximum for
  $B\omega_{\mathrm{rms}}\sim \mathcal{O}(1)$ is understood noticing
  that $-\langle \mathrm{p}_y \omega_x\rangle$ must decrease for
  $B\omega_{\mathrm{rms}} \gg 1$.  In the limit
  $B\omega_{\mathrm{rms}}\!\! \ll 1$, $-\langle
  \mathrm{p}_y\omega_x\rangle\simeq B\omega^2_{\mathrm{rms}}$ (solid
  line) is implied by $\mathrm{\bf p}\simeq (B\omega_x,-B\omega_y,1)$,
  see text. In the random tumbling limit $\langle \mathrm{p}_x^2 \rangle =
  \langle p^2_y\rangle \to 1/3$ (Fig.~\ref{fig3}a), which implies
  $-\langle \mathrm{p}_y\omega_x\rangle=\langle p^2_y\rangle_s/B\sim
  1/(3B)$ (dashed line).}
\label{fig4}
\end{figure}
We now consider the limit $a_{\mathrm{rms}} \ll\ g$ 
when we can take $\bm A=-\bm g$ and Eq.~(\ref{eq:2}) reads
\begin{equation}
\dot{{\mathrm{\bf p}}} = \frac{1}{2B}(\hat{\bf z} - \mathrm{p}_z {\mathrm{\bf p}}) +\frac{1}{2} {\bm \omega} \times {\mathrm{\bf p}}\, ,
\label{eq:5}
\end{equation} 
with $B\!=\!v_o/g$. Similarly to the previous case, when
$B\omega_{\mathrm{rms}} \to 0$ the cells orient in the preferred
direction $\hat{\bm z}$, which is now fixed in space.
The effective velocity thus becomes $\bm
v=\bm u+v_s \hat{\bm z}$ which, unlike the previous case, is
incompressible (${\bm \nabla} \cdot {\bm v}=0$).  
Therefore, now we expect that not only for 
$B\omega_{\mathrm{rms}}\gg 1$ but also for
$B\omega_{\mathrm{rms}} \to 0$ swimmers distribute uniformly, as
confirmed by Fig.~\ref{fig3}a showing that $D_2 \to 3$ in both
limits. Remarkably, Fig.~\ref{fig3}a shows that also in this case gyrotactic
swimmers cluster on a fractal set (see Fig.~\ref{fig1}b)
for intermediate values, with a well defined minimum of the 
correlation dimension ($D_2\approx 2.7$)
for $B\omega_{\mathrm{rms}} \sim \mathcal{O}(1)$. We remark than an
optimal orientation timescale for aggregation is also observed in
steady kinematic vortical flows \cite{Durham2011} where, however, a
vast class of trajectories is integrable.

We can understand the origin of the observed clustering by considering
the limit $B\omega_{\mathrm{rms}}\ll 1$. In such limit, cell
orientation being very fast we can assume that the swimming direction
$\mathrm{\bf p}$ is always at an equilibrium orientation with $\mathrm{p}_x,
\mathrm{p}_y \ll \mathrm{p}_z \simeq 1$ (see Fig.~\ref{fig3}b). 
In particular, solving Eq.~(\ref{eq:5}) for
$\dot{\mathrm{\bf p}}=0$, at first order in $\mathrm{p}_x,\mathrm{p}_y$, 
one finds
$\mathrm{p}_x \simeq B \omega_y$ and $\mathrm{p}_y \simeq - B
\omega_x$ (which is confirmed by simulations). As a consequence, the
effective swimmer velocity field ${\bm v}={\bm u} + v_s \mathrm{\bf
p}$ with $\mathrm{\bf p}\simeq (B\omega_y,-B\omega_x,1)$  
has a compressible component with divergence
\begin{equation}
{\bm \nabla} \cdot {\bm v}\simeq - v_sB \nabla^2 u_z \, ,
\label{eq:6}
\end{equation} 
which, unlike the previous case, is unrelated to fluid acceleration so
that swimmers will cluster in regions different from those of high
vorticity (compare Fig.~\ref{fig1}a and b).  We notice that
(\ref{eq:6}) generalizes the well known mechanism of cell focusing in
the center (walls) of downward (upward) vertical pipe flows
\cite{Kessler1985}.  Notice that in the above argument the vertical
component of the vorticity plays no role, as it does not change
$\mathrm{p}_z$.

Another consequence of the expansion $\mathrm{\bf p}\simeq
(B\omega_x,-B\omega_y,1)$ is that $\mathrm{p}_{x}$
(resp. $\mathrm{p}_y$) and $\omega_{y}$ ($\omega_{x}$) have locally
the same (opposite) sign.  Numerical simulations show that this
remains true also for larger values of $B\omega_{\mathrm{rms}}$, on
average. Indeed, at stationarity, by averaging Eq.~(\ref{eq:5}) and
using isotropy on the $(x,y)$ plane (guaranteed by the isotropy of the
fluid velocity field) we obtain $\langle \mathrm{p}^2_x\rangle=\langle
\mathrm{p}^2_y\rangle= B\langle \mathrm{p}_x\omega_y\rangle= -B\langle
\mathrm{p}_y\omega_x\rangle$.  The correlation between the horizontal
components of $\mathrm{\bf p}$ and $\bm\omega$ implies that the
swimmers will stay longer in regions of the flow characterized by
positive vertical velocity and negative vertical acceleration
(Fig.~\ref{fig4}a). This can be easily seen in a case with, say, a
vortex aligned with the x-axis, where the above argument with
$\omega_x>0$ implies $\langle \rm \mathrm{p}_z\!\rangle>\!0$, $\langle
\rm \mathrm{p}_y\!\rangle\!<0$, so that the trajectories spend more
time in regions where $a_z\!>\!0$,$u_z\!<\!0$ as there the swimming
velocity opposes that of the fluid.  The preferential concentration in
these regions of the flow will be maximal (and correspondingly the
correlation dimension minimal, i.e. clustering stronger) for
$B\omega_{\mathrm{rms}}\sim \mathcal{O}(1)$ where the correlation
between swimming direction and vorticity $-\langle
\mathrm{p}_y\omega_x\rangle=\langle \mathrm{p}_x\omega_y\rangle$ is
also maximal (Fig.~\ref{fig4}b). For such value of
$B\omega_{\mathrm{rms}}$ fluid regions with maximal deviation of the
swimming direction from the vertical will balance vorticity dominated
ones where $\mathrm{\bf p}$ tumbles randomly. We observe that the
swimmer vertical migration can be strongly inhibited by the bias
towards downwelling regions: in Fig.~\ref{fig4}a, e.g., $\langle u_z
\rangle$ can reach $30\%$ of the swimming speed $v_s$.

\begin{figure}[t!]
\centering
\includegraphics[width=0.95\columnwidth]{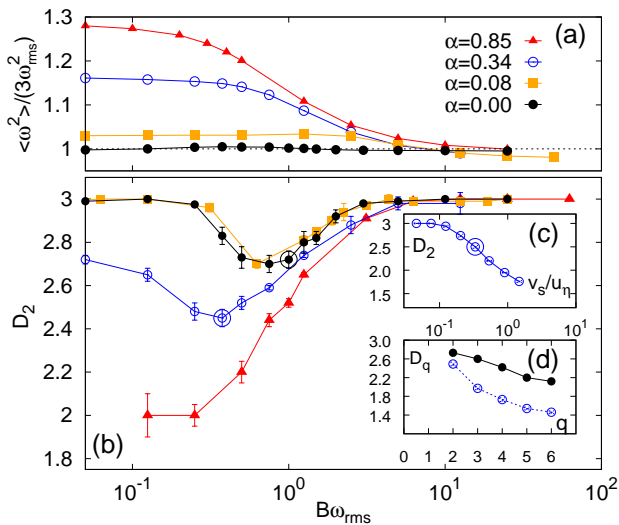}
\caption{(color online) (a) Average square vorticity at swimmers
  position normalized to the volume average value at varying the ratio
  $\alpha=a_{\mathrm{rms}}/g$ ($\alpha=0$ corresponds to data of
  Fig.~\ref{fig3}); (b) correlation dimension $D_2$ vs $\alpha$; (c)
  $D_2$ vs swimming speed $v_s/u_{\eta}$, the circled symbol
  corresponds to the circled one in (b); (d) generalized dimensions
  $D_q$ vs $q$ for circled data in (b), notice that the case
  $\alpha=0$ (filled black circles) appears to be less multifractal
  than when also the fluid acceleration is contributing to clustering
  (empty blue circles). }
\label{fig5}
\end{figure}
In the general case, the relative importance of fluid and
gravitational accelerations for clustering depends on the ratio
$\alpha=a_{\mathrm{rms}}/g$.  Figure~\ref{fig5}a indeed shows that the
bias towards regions of high vorticity decreases with $\alpha$ and is
absent when only the gravitational torque is acting ($\alpha=0$). The
correlation dimension $D_2$, shown in Fig.~\ref{fig5}b, smoothly
varies with $\alpha$, interpolating from the two limits shown in
Fig.~\ref{fig2}c and \ref{fig3}b.  We observe that, as anticipated,
clustering is more effective for large swimming speeds as displayed in
Fig.~\ref{fig5}c, showing that, at fixed value of 
$B \omega_{\mathrm{rms}}$, $D_2$ decreases with $v_s/u_\eta$.
Finally, as one can expect from general considerations on dynamical
attractors \cite{Bec2005}, Fig.~\ref{fig5}d demonstrates that the
spatial distribution of the gyrotactic self-propelled particles is
multifractal, as the generalized dimensions $D_q$ (controlling the
probability to find $q$ particles at small separation) depends on the
moment $q$ \cite{paladin1987}.

Summarizing, we have shown that gyrotactic motility and realistic
turbulent flows can generate small-scale patchiness (down to the
Kolmogorov scale) in the distribution of bottom-heavy swimming
microorganisms. We identified two mechanisms driving microorganism
clustering: the focusing in vortical regions due to local adjustment
of the swimming orientation with fluid acceleration, and the
correlation between vorticity and swimming direction on the plane
perpendicular to gravity leading particles to preferentially explore
downwelling, upward accelerating regions. In general, gravity is
expected to dominate when turbulent intensity is not very high and it
is likely the most important effect in the ocean.  Crucial parameters
for observing clustering are in this case the ratio between swimming
speed and small-scale fluid velocity fluctuations ($v_s/u_\eta$) and
the reorientation time scale with respect to vorticity intensity
($B\omega_{\mathrm{rms}}$). For typical microalgae $B\approx 1\!-\!6s$
and $v_s=100\!-\!200\mu m/s$ \cite{Jones94,hill97,Kessler1985}. In
the ocean, the turbulence intensity, measured in terms of kinetic energy
dissipation $\epsilon$, varies from $\epsilon \sim
10^{-4}\!-\!10^{-5}\,W/Kg$ in the upper mixing layer down to $\epsilon
\sim 10^{-6}\!-\!10^{-7}\,W/Kg$ a few meters deeper
\cite{Yamazaki1996,Yamazaki2002}. We can thus estimate that
$v_s/u_\eta \in [0.02\!:\!0.4]$ and $B\omega_{\mathrm{rms}} \in
[0.1\!:\!50]$ therefore the effects discussed in this Letter are
relevant in realistic conditions and can definitely be tested in
laboratory by tuning turbulence characteristics.

We conclude by remarking that for non-spherical cells such as, e.g.,
prolate spheroids the term $\gamma \mathrm{\bf p} \cdot
\mathbb{S}\cdot(\mathbb{I}-\mathrm{\bf p}\otimes\mathrm{\bf p})$
should be added to Eq.~(\ref{eq:2}) ($\gamma$ being the eccentricity,
and $\mathbb{S}$ and $\mathbb{I}$ the symmetric rate of strain tensor
and identity matrix, resp.)  \cite{Pedley1987}. Such term is also
contributing to the phase-space contraction rate (\ref{eq:4})
providing an additional mechanism for clustering \cite{Torney2007}. It
will thus be interesting to study if and how gyrotactic clustering in
turbulence is modified at varying the cell shape. 

\begin{acknowledgments}
We thanks S. Musacchio for useful discussions.  GB and MC acknowledge
KITPC institute for hospitality during the program \textit{New
  Directions in Turbulence} and support by MIUR PRIN-2009PYYZM5
``Fluttuazioni: dai sistemi macroscopici alle nanoscale''.
\end{acknowledgments}

%

\end{document}